\documentclass[twocolumn,showpacs,preprintnumbers,amsmath,amssymb,prl]{revtex4}

\usepackage{graphicx}
\usepackage{dcolumn}
\usepackage{bm}

\begin{document}

\title{The percolation transition of hydration water: from planar hydrophilic surfaces to proteins}

\author{Alla Oleinikova}
\email{alla@pc2a.chemie.uni-dortmund.de} 
\author{Ivan Brovchenko, Nikolai Smolin, Aliaksei Krukau, Alfons Geiger}
\author{Roland Winter}%
\affiliation{Physical Chemistry, University of Dortmund, Otto-Hahn-Str.6, Dortmund, D-44227, Germany
}

\date{\today}
\begin{abstract}
The formation of a spanning hydrogen-bonded network of hydration water is
found to occur via a 2D percolation transition in various systems: smooth hydrophilic surfaces, the surface of a single protein molecule, protein powder and diluted peptide solution. The average number of water-water hydrogen bonds \textit{n$_H$} at the percolation threshold varies from 2.0 to 2.3, depending on temperature, system size and surface properties. Calculation of \textit{n$_H$} allows an easy estimation of the percolation threshold of hydration water in various systems, including biomolecules.  
\end{abstract}

\maketitle
The existence of a spanning network of hydration water in biosystems enables their
biological functions \cite{Careri2,Rupley,Bruni,Careri,Sok,Care3}. With increasing hydration level, an ensemble of finite (non-spanning) clusters of hydration water transforms via a quasi-2D percolation transition into a state with a spanning water network  \cite{Careri,lys1}. The full dynamics of biomolecules is restored, when they are covered by about a "monolayer" of water. The first appearance of such a "monolayer" corresponds to a quasi-2D percolation transition of the hydration water at the surface of a single biomolecule \cite{lys1,lys2}. Simulation studies of various properties of hydrated biosystems below and above the percolation threshold of the hydration water can help to clarify the role of the spanning water network in the onset of biological functions. Such studies require the knowledge of the percolation threshold of water in the system of consideration. This information can be obtained by conventional percolation analysis of water clustering \cite{Stauffer,Geiger2,THF,THF2}, which is extremely time consuming. In the present paper we propose a simple method to locate the percolation threshold of hydration water even in complex systems, using the average number of water-water hydrogen bonds. This method is derived from extensive computer simulation studies of the percolation transition of hydration water in various systems: water adsorbed at smooth hydrophilic planes and spheres \cite{lys1,handbook,plane}, at surfaces of rigid and flexible single lysozyme molecules and lysozyme powder \cite{lys1,lys2}, hydration water in protein solutions \cite{Alex}. 
\par
Water molecules are considered to belong to the same cluster if they are connected by an uninterrupted path of hydrogen-bonds. Water clustering was studied by computer simulations at various hydration levels in low-hydrated systems \cite{lys1,plane,handbook} and at various temperatures and widths of the hydration shell for proteins in solution \cite{Alex}. The percolation threshold of water at planar surfaces and in a model lysozyme powder (both systems are infinite due to the periodic boundary conditions), can be located by using the spanning probability, cluster size distribution and fractal dimension of the largest cluster. In closed systems, such as a surface of a finite object, an \textit{infinite} cluster can not appear and the spanning probability is not defined. However, the formation of a \textit{spanning} network of hydration water at the surface of a smooth hydrophilic sphere and of a single biomolecule, was found to occur in a way similar to infinite systems \cite{lys1,lys2,handbook,Alex}. 
\begin{figure}[htb]
\includegraphics[width=7cm]{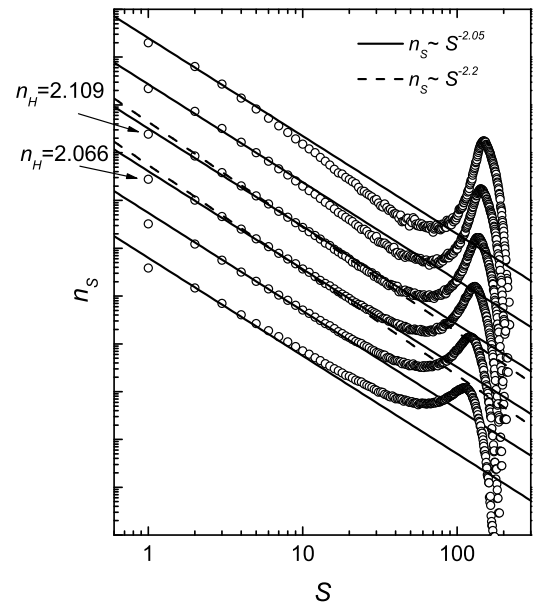}
\caption{Cluster size distribution \textit{n$_S$} in the hydration shell of an elastin-like peptide at \textit{T} = 300 K for several widths \textit{D} of the hydration shell, which corresponds to the following numbers \textit{N$_w$} of water molecules in the shell: 135, 141, 147, 153, 159 and 164 (from bottom to top). Power laws for 2D and 3D percolation thresholds are shown by solid and dashed lines, respectively.}
\end{figure}
\par
The probability distribution \textit{n$_S$} of cluster sizes \textit{S} obeys a universal power law \textit{n$_S$} $\sim$ \textit{S}$^{-\tau}$ in a widest range of \textit{S} at the percolation threshold, with $\tau$ $\approx$ 2.05 for 2D and $\tau$ $\approx$ 2.2 for 3D percolation \cite{Stauffer}. In Fig. 1 we show the cluster size distributions \textit{n$_S$} of water molecules in the hydration shell of the smallest studied system, the fully hydrated elastin-like peptide GVG(VPGVG)$_3$, for various sizes of the shell width \textit{D}. A hump in \textit{n$_S$} at large \textit{S} reflects the truncation of the large clusters due to the finite size of the hydration shell. The behavior of \textit{n$_S$} allows the location of the percolation threshold between \textit{N$_w$} = 147 and \textit{N$_w$} = 153, even without any assumption about the dimensionality of the transition.
\par
The dimensionality of the percolation transition can be determined from the effective fractal dimension \textit{d$_f$} of the largest water cluster, which can be obtained from a fit of its mass distribution: \textit{m(r) $\sim$ r~$^{ d_f}$}. The range of \textit{r}, used in the fits of \textit{m(r)}, is determined by the system size. We limited this range by the half of the simulation box \textit{L} for planar surfaces and lysozyme powder, by the diameter of the spherical surfaces and by the shortest value of the effective axis of a single lysozyme molecule, which is about 24 $\mbox{\AA}$. At the percolation threshold, the fractal dimension of the largest cluster is \textit{d$_f^{2D}$} $\approx$ 1.9 for a 2D system and \textit{d$_f^{3D}$} $\approx$  2.53 for a 3D system \cite{Stauffer}. We have found that in low-hydrated systems the value of \textit{d$_f$} is close to the value \textit{d$_f^{2D}$} \cite{lys1,handbook,plane}. This indicates the quasi-2D character of the percolation of adsorbed water. In particular, the threshold hydration level and the dimensionality of the water percolation transition in model lysozyme powder remarkably agrees with experimental results \cite{Careri}. Note that the fractal dimension of the largest water cluster in the hydration shell of the elastin-like peptide in solution can not be determined due to the small system size (the average radius of gyration of this peptide does not exceed about 8 $\mbox{\AA}$).
\begin{table}[htb]
\caption{Number of water molecules in the first (\textit{N$_1$}) and second
  (\textit{N$_2$}) hydration shell per unit surface area and the average 
number of water-water hydrogen-bonds \textit{n$_H$} at the quasi-2D percolation 
thresholds of various model systems and proteins.  }
\label{tab:par} 
\vspace{0.5cm}     
\begin{tabular}{c|c|c|c|c}
\hline\noalign{\smallskip}
System & \textit{T} / K & \textit{N$_1$} / 100 $\mbox{\AA}^2$ &\textit{N$_2$} 
/ \textit{N$_1$} & \textit{n$_H$} \\
 &  & ($\pm$ 0.05) & ($\pm$ 0.005) & ($\pm$ 0.01)\\
\noalign{\smallskip}\hline\noalign{\smallskip}
Plane 80x80 $\mbox{\AA}^2$ & 425 & 7.5 & 0.043 & 2.20 \\
Plane 100x100 $\mbox{\AA}^2$ & 425 & 7.4 & 0.043 & 2.22 \\
Plane 150x150 $\mbox{\AA}^2$ & 425 & 7.3 & 0.044 & 2.22 \\
Sphere \textit{R} = 15 $\mbox{\AA}$ & 425 & 8.6 & 0.12 & 2.11 \\
Sphere \textit{R} = 30 $\mbox{\AA}$ & 425 & 8.5 & 0.11 & 2.14 \\
Sphere \textit{R} = 50 $\mbox{\AA}$ & 425 & 8.3 & 0.10 & 2.15 \\
Sphere \textit{R} = 15 $\mbox{\AA}$ & 475 & 9.1 & 0.36 & 1.95 \\
Lysozyme powder & 300 & 2.1\footnotemark[1] & - & 2.32 \\
Rigid lysozyme & 300 & 4.9 & 0.38 & 2.30 \\
Flexible lysozyme & 300 & 5.0 & 0.44 & 2.30 \\
Lysozyme powder & 400 & 2.5\footnotemark[1] & - & 2.02 \\
Rigid lysozyme & 400 & 5.1 & 0.98 & 2.05 \\
Elastin & 320 & 10.8 & - & 2.08 \\
Elastin & 300 & 10.3 & - & 2.10 \\
Elastin & 280 & 9.8 & - & 2.10 \\
Elastin & 260 & 9.2 & - & 2.08 \\
\noalign{\smallskip}\hline
\end{tabular}
\footnotetext[1]{total number of water molecules per one lysozyme}
\end{table} 
\par
The percolation thresholds of hydration water in all systems studied are given in Table I in terms of \textit{N$_1$}, the number of water molecules in the first hydration shell per
unit surface area. For lysozyme powder, \textit{N$_1$} is simply the total number of water molecules in the simulated system divided by the number of protein molecules.
The values obtained for the threshold hydration level in terms of \textit{N$_1$}
noticeably vary from one studied system to another and also depend on
temperature. 
These variations of \textit{N$_1$} at the threshold should be expected, 
as this surface coverage is similar to the occupancy variable in lattice models, which is essentially non-universal. For random percolation in 2D lattices, the threshold value of site and bond occupancy depends on the lattice type \cite{Stauffer}. The honeycomb lattice with 3 neighbors and the square lattice with 4 neighbors seem to be the most relevant for water adsorbed at hydrophilic surfaces \cite{jml02,handbook}. For these lattices, the threshold values of the occupancy variable are $\sim$ 0.70 and $\sim$ 0.59 for site percolation and $\sim$ 0.65 and 0.50 for bond percolation, respectively. These threshold values are more universal in terms of the average number of bonds, which is $\sim$ 2.09 and $\sim$ 2.37 for site percolation and $\sim$ 1.96 and 2.00 for bond percolation, on the honeycomb and square lattice, respectively. As the water percolation process can be considered as a correlated site-bond percolation \cite{ST}, one may expect a rather universal threshold value in various hydrated systems in terms of \textit{n$_H$}, the average number of water-water H-bonds, formed by each water molecule. Such expectation is supported by the observed similarity between percolation in bulk liquid water and random bond percolation in 3D lattice. In bulk liquid water, \textit{n$_H$} at the percolation threshold is about 1.53 \cite{Geiger}, i.e., rather close to the value 1.55 for bond percolation in a 3D diamond lattice with 4 neighbors \cite{Stauffer}. 
\begin{figure}[htb]
\includegraphics[width=8cm]{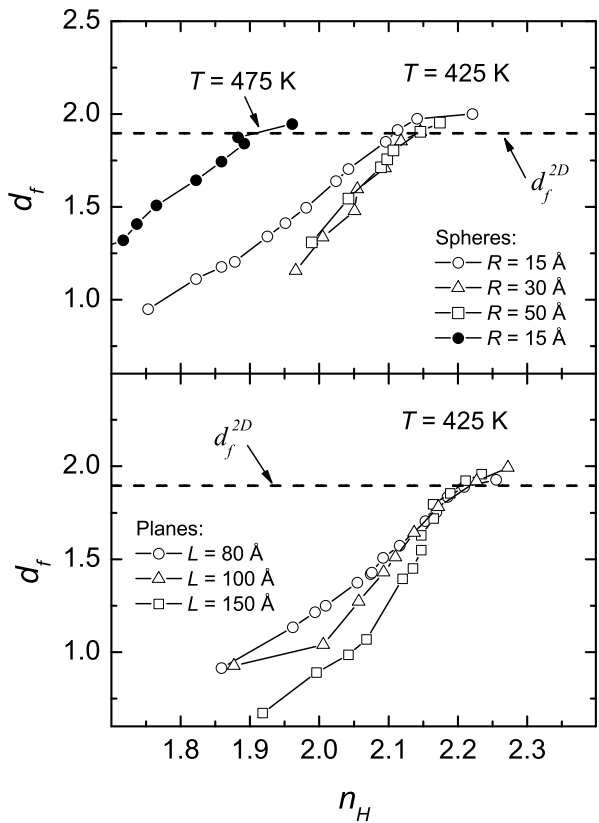}
\caption{\label{fig:plane} Fractal dimension of the largest cluster \textit{d$_f$} as a function of the average number \textit{n$_H$} of H-bonds between water molecules near smooth hydrophilic planar (lower panel) and spherical (upper panel) surfaces. }
\end{figure}
\par
The dependence of the fractal dimension \textit{d$_f$} on \textit{n$_H$} for some of the systems studied is shown in Figs. 2 and 3. Below the percolation threshold, \textit{d$_f$} is
essentially an \textit{effective} fractal dimension, because most of the largest water
clusters are not fractal objects. That is why the values of \textit{d$_f$} noticeably depend on the system size and geometry at low hydration levels. At the percolation threshold, the structure of the largest water cluster is close to a fractal and \textit{d$_f$} approaches the threshold fractal dimension \textit{d$_f^{2D}$} for 2D percolation. For three studied planar surfaces of different sizes, the percolation thresholds of hydration water practically coincide: at \textit{T} = 425 K they are located at \textit{n$_H$} = 2.21$\pm$0.01 and \textit{N$_1$} = 0.074$\pm$0.001 $\mbox{\AA}^{-2}$. At the spherical surfaces, the percolation threshold corresponds to \textit{N$_1$} = 0.085$\pm$0.002 $\mbox{\AA}^{-2}$, that is about 15$\%$ higher than the threshold value of \textit{N$_1$} for planar surfaces. However, the value of \textit{n$_H$} at the threshold is 2.13$\pm$0.02 for spherical surfaces, that is just 4$\%$ lower than for the planar surface. The higher threshold hydration levels \textit{N$_1$} for the spheres are accompanied by higher concentrations of water molecules in the second hydration shell \textit{N$_2$}. The ratio \textit{N$_2$}/\textit{N$_1$} is almost 12$\%$ for spheres, whereas it is only 4$\%$ for the planar surfaces of the
same hydrophilicity (see Table I). The presence of water molecules in the second shell effectively reduces the number of water-water H-bonds in the first shell, thus resulting in a higher threshold value of \textit{N$_1$} and a lower threshold value of \textit{n$_H$}. Such shift of \textit{n$_H$} to lower values can be considered as a trend toward three-dimensionality, where \textit{n$_H$} is about 1.53 at the percolation threshold of bulk liquid water \cite{Geiger}. 
\begin{figure}[htb]
\includegraphics[width=8cm]{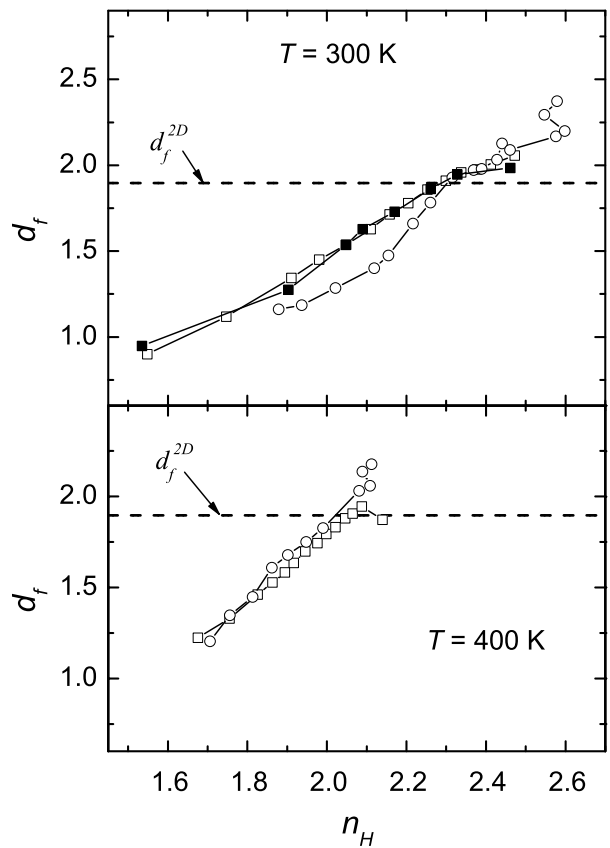}
\caption{\label{fig:lysosyme} Fractal dimension of the largest cluster \textit{d$_f$} as a function of the average number \textit{n$_H$} of H-bonds between water molecules at the surface of rigid (open squares) and flexible (solid squares) lysozyme and in the hydrated lysozyme powder (open circles). }
\end{figure}
\par
A similar trend toward larger threshold values of \textit{N$_1$} and smaller values of \textit{n$_H$} is observed with increasing temperature. When the temperature rises from \textit{T} = 425 to 475 K, the threshold value of \textit{N$_1$} for a sphere with radius \textit{R} = 15 $\mbox{\AA}$ increases from 0.086 to 0.091$\mbox{\AA}^{-2}$, whereas the fraction of molecules in the second hydration shell increases and \textit{N$_2$}/\textit{N$_1$} changes from 0.12 to 0.36. This is accompanied by a decrease of \textit{n$_H$} from 2.11 to 1.95. Hence, an increase of temperature by 100 degrees causes a decrease of the \textit{n$_H$} value at the threshold by about 15 $\%$. This trend corresponds to the growing importance of the "bond" percolation relatively to the "site" percolation with increasing temperature in site-bond percolation of water.   
\par
At the surfaces of the rigid lysozyme molecule, the flexible lysozyme molecule, and in the model lysozyme powder, the percolation transition of hydration water at ambient conditions (\textit{T} = 300 K) occurs when the average number of water-water H-bonds \textit{n$_H$} is 2.31$\pm$0.01 (see Table I and Fig. 3). Intriguingly, the value of \textit{n$_H$} at the percolation threshold remains highly universal even in such complex systems as hydrated proteins. An increase of the temperature to \textit{T} = 400 K reduces the threshold value of \textit{n$_H$} to 2.03$\pm$0.02, i.e by about 15 $\%$. Hence, the effect of temperature on the threshold value of \textit{n$_H$} is remarkably similar at protein surfaces and at smooth surfaces of essentially hydrophilic spheres (compare the values \textit{N$_2$}/\textit{N$_1$} for these systems in Table I). 
\par
The surface of a biomolecule in dilute aqueous solution is completely covered by hydration water, which forms a spanning network at low temperature. Upon heating, this spanning water network breaks up via a percolation transition, and the hydration water shell becomes an ensemble of finite (non-spanning) water clusters \cite{Alex}. The location of the percolation threshold at the surface of fully hydrated elastin-like peptide, using cluster size distribution \textit{n$_S$}, indicates, that at \textit{T} = 300 K, \textit{n$_H$} is about 2.09 $\pm$ 0.02 at the threshold (see Fig. 1). More accurately, the percolation threshold can be located by additional analysis of the probability distribution of the largest cluster size \cite{plane,Alex}. The estimated temperature of the percolation transitions of hydration water in the shells of various width \textit{D} are shown by open circles in Fig. 4 as a function of the average value of water-water H-bonds \textit{n$_H$}. Evidently, the value of \textit{n$_H$} at the percolation threshold is about 2.1 for any reasonable choice of the hydration shell width \textit{D}. The value \textit{n$_H$} for hydration water in protein solutions decreases almost linearly with temperature (see Fig. 4 and also Ref. \cite{temp}). This gives us the possibility to estimate the value of \textit{n$_H$} in a wide temperature range by simulations at two temperatures and, subsequently, to estimate the temperature of the thermal breaking of the spanning network of hydration water. 
\begin{figure}[htb]
\includegraphics[width=8cm]{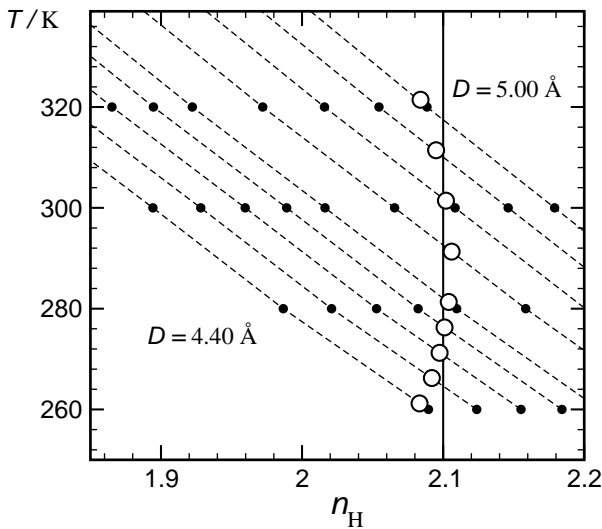}
\caption{\label{fig:elastin} The average value \textit{n$_H$} of water-water H-bonds  in the hydration shell of an elastin-like peptide as a function of temperature (closed circles and dashed lines). The widths of the hydration shell \textit{D} are: 4.40, 4.45, 4.50, 4.55, 4.60, 4.70, 4.80, 4.90, and 5.00 $\mbox{\AA}$. Percolation thresholds are shown by open circles.}
\end{figure}
\par   
Summarizing our studies of the quasi-2D percolation transition of hydration water in various model systems, we propose a relatively simple method to indicate the existence of a spanning network of hydration water by the analysis of the average number \textit{n$_H$} of hydrogen bonds between the water molecules in the hydration shell. This value can be obtained by conventional computer simulations or estimated from experimental data \cite{Nak}. At ambient temperature, the threshold value of \textit{n$_H$} is about 2.1 for fully hydrated systems and about 2.3 for low-hydrated systems. A value \textit{n$_H$} above the threshold indicates the presence of a spanning network of hydration water. The threshold value of \textit{n$_H$} decreases slightly with increasing temperature. 
\par
We thank G. Careri for stimulating discussions and the Deutsche Forschungsgemeinschaft (DFG-Forschergruppe 436) for financial support. A.K. thanks the International Max Planck Research School in Chemical Biology for financial support.
\bibliography{percprlrev1}
\end{document}